\documentclass[reprint,aps,prl,superscriptaddress,amsmath,amssymb,floatfix]{revtex4-1}
\usepackage{graphicx}
\usepackage{layout}
\usepackage{natbib}
\usepackage{dcolumn}
\usepackage{braket}
\usepackage{bm}
\usepackage{color}
\usepackage{verbatim}
\usepackage{hyperref}

\begin{document}

\title{A Parity-Protected Superconductor-Semiconductor Qubit}

\author{T.~W.~Larsen}
\affiliation{Center for Quantum Devices and Microsoft Quantum Lab--Copenhagen, Niels Bohr Institute, University of Copenhagen, 2100 Copenhagen, Denmark}

\author{M.~E.~Gershenson}
\affiliation{Department of Physics and Astronomy, Rutgers University, Piscataway, New Jersey 08854, USA}

\author{L.~Casparis}
\affiliation{Center for Quantum Devices and Microsoft Quantum Lab--Copenhagen, Niels Bohr Institute, University of Copenhagen, 2100 Copenhagen, Denmark}

\author{A.~Kringh\o j}
\affiliation{Center for Quantum Devices and Microsoft Quantum Lab--Copenhagen, Niels Bohr Institute, University of Copenhagen, 2100 Copenhagen, Denmark}

\author{N.~J.~Pearson}
\affiliation{Center for Quantum Devices and Microsoft Quantum Lab--Copenhagen, Niels Bohr Institute, University of Copenhagen, 2100 Copenhagen, Denmark}
\affiliation{Theoretische Physik, ETH Zurich, 8093 Zurich, Switzerland}

\author{R.~P.~G.~McNeil}
\affiliation{Center for Quantum Devices and Microsoft Quantum Lab--Copenhagen, Niels Bohr Institute, University of Copenhagen, 2100 Copenhagen, Denmark}

\author{F.~Kuemmeth}
\affiliation{Center for Quantum Devices and Microsoft Quantum Lab--Copenhagen, Niels Bohr Institute, University of Copenhagen, 2100 Copenhagen, Denmark}

\author{P.~Krogstrup}
\affiliation{Center for Quantum Devices and Microsoft Quantum Lab--Copenhagen, Niels Bohr Institute, University of Copenhagen, 2100 Copenhagen, Denmark}
\affiliation{Microsoft Quantum Materials Lab--Copenhagen, 2800 Kongens Lyngby, Denmark}

\author{K.~D.~Petersson}
\affiliation{Center for Quantum Devices and Microsoft Quantum Lab--Copenhagen, Niels Bohr Institute, University of Copenhagen, 2100 Copenhagen, Denmark}

\author{C.~M.~Marcus}
\affiliation{Center for Quantum Devices and Microsoft Quantum Lab--Copenhagen, Niels Bohr Institute, University of Copenhagen, 2100 Copenhagen, Denmark}

%\date{\today}

\begin{abstract}
Coherence of superconducting qubits can be improved by implementing  designs that protect the parity of Cooper pairs on superconducting islands. Here, we introduce a parity-protected qubit based on voltage-controlled semiconductor nanowire Josephson junctions, taking advantage of the higher harmonic content in the energy-phase relation of few-channel junctions. A symmetric interferometer formed by two such junctions, gate-tuned into balance and frustrated by a half-quantum of applied flux, yields a $\cos(2\varphi)$ Josephson element, reflecting coherent transport of pairs of Cooper pairs. We demonstrate that relaxation of the qubit can be suppressed ten-fold by tuning into the protected regime.
\end{abstract}

\pacs{}

\maketitle

%% Introduction
Recent proof-of-concept demonstrations of quantum simulations have highlighted progress in the development of small-scale quantum processors \cite{OMalley:2016fh,Kandal:2017aa}. While potentially useful for near-term applications \cite{preskill_2018,Arute:2019aa}, the qubits used in the experiments were still susceptible to errors, limiting the accuracy and complexity of quantum algorithms that these systems can support. Ultimately, qubits with fault-tolerant operations are desired \cite{DiVinc:2000aa}. Ideas for fault-tolerant quantum computers rely on redundantly encoding quantum information in a protected subspace of a larger quantum system \cite{Terhal:2015ks}. One approach is through the use of quantum error correction codes such as surface or color codes, which actively perform stabilizer measurements to confine a large Hilbert space to a subspace that is protected from local, random errors \cite{Knill:2005eu}. Quantum error correction is expected to allow fault-tolerant quantum computing using noisy qubits at the cost of requiring many physical qubits for each logical qubit and increased runtime \cite{Fowler:2012fi,Reiher:2016aa}.

An alternative approach is to engineer fault tolerance at the device level. This can be implemented, for instance, using Majorana zero modes in a network of topological superconductors \cite{Kitaev:2001kla,Kitaev:1997wr}, forming  a highly degenerate ground state in which quantum information can be encoded nonlocally, protecting it from local noise. Another form of device-level protection can be implemented using Josephson junctions (JJs) with potentials that are $\pi$-periodic in the phase difference, $\varphi$, of the superconducting order parameter across the junction \cite{Doucot:2012kt}. Similar to Majorana qubits, $\pi$-periodic JJs protect quantum information using disconnected parity subspaces. For Majoranas, it is the parity of the number of electrons on an island that is relevant; for $\pi$-periodic JJs it is the parity of the number of Cooper pairs on an island. Protected $\pi$-periodic JJs also allow protected quantum operations \cite{Brooks:2013hi}, suggesting a fruitful path towards fault-tolerant quantum computing.

Several implementations of $\pi$-periodic Josephson devices have appeared \cite{Kitaev:2006ud,Gyenis:2019}. A recent version \cite{Bell:2014de} used four JJs in a rhombus configuration to generate a $\pi$-periodic $\cos\,(2\varphi)$ potential, yielding a qubit defined by the parity of Cooper pairs. This qubit is dual to the the recently introduced bifluxon qubit, defined by the parity of flux quanta \cite{kalashnikov:2019bg}. The Josephson rhombus uses four nominally identical JJs in a loop \cite{Doucot:2002et, Gladchenko:2009dj}. Departures from symmetry, for instance due to fabrication variation among the four junctions, lifts the degeneracy of the lowest two states and reduces protection. 

Here, we implement the $\cos\,(2\varphi)$ element needed for protection using a pair of gate-tunable semiconductor JJs based on InAs nanowires grown with epitaxial superconducting Al \cite{Krogstrup:2015en,vanWoerkom:2017ek,goffman_2017}. Nanowire-based superconducting qubits, or gatemons \cite{Larsen:2015cp, luthi_2018} and two-junction superconducting quantum interference devices (SQUIDs) \cite{deLange:2015gv} have been explored recently. A two-gatemon SQUID is particularly useful for creating protected qubits, as gate-control of junction transmission allows precise {\it in-situ} balancing of the interferometer at fixed external flux, and, critically, makes use of higher harmonics of the energy-phase relation for a few-channel semiconductor junction \cite{spanton_2017, Kringhoj:2017uh, Hays:2017ud} to create a robust and tunable $\pi$-periodic qubit. We observe that when the interferometer is gate-tuned {\it in situ} into balance, the resulting protected qubit shows a factor-of-ten enhancement in lifetime compared to unprotected tunings.

\begin{figure}
\includegraphics[width=1\columnwidth]{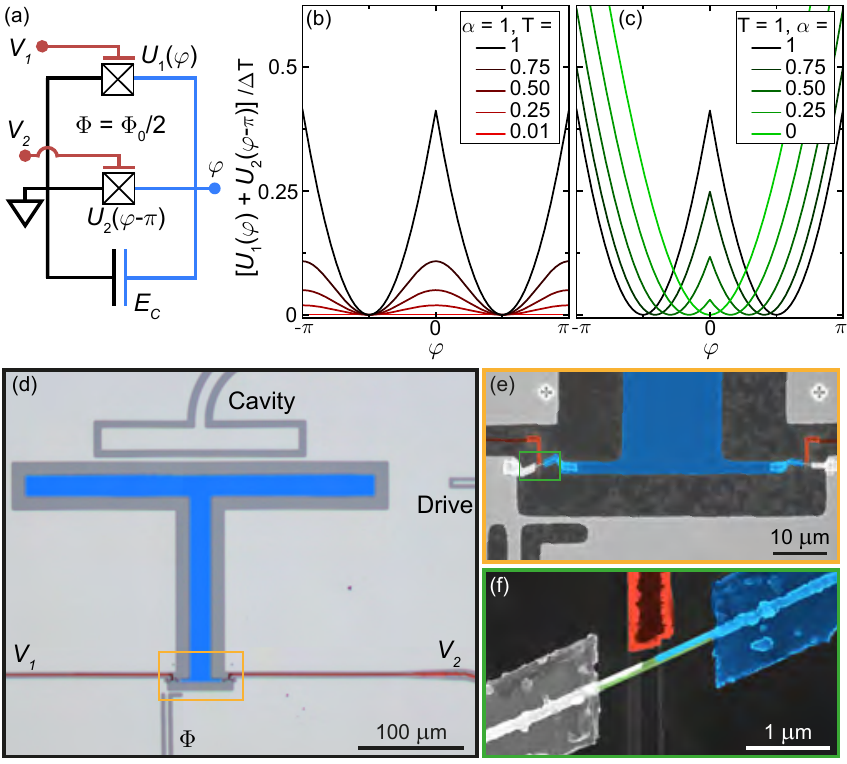}
\caption{\label{fig:1}
Qubit circuit and device design. (a) Circuit schematic of the parity-protected qubit formed from high-transparency few-channel semiconductor Josephson junctions in a flux-biased interferometer shunted by a  large capacitor (blue). (b, c) Energy-phase relation of the interferometer for (b) different junction transmission coefficients and (c) junction asymmetries. (d) False-color optical micrograph of the device showing the large island (blue) that forms one side of the shunting capacitor. (e, f) False-color electron micrographs of the nanowire junctions. (f) A small segment of the Al shell on an InAs nanowire is etched away to form a semiconductor Josephson junction. A nearby electrostatic gate (red) allows tuning of the electron density in the junction. 
}
\end{figure}

%% Theory
The protected qubit circuit is shown in Fig.~1(a). The transmon-like geometry consists of a superconducting island with charging energy $E_C$ connected to ground through two JJs in a SQUID configuration. Junction transmissions are tuned using gate voltages $V_k$ ($k=1,2$). We model the two JJs in the short-junction regime, expressing Josephson coupling as mediated by a number ($i=1,2, ...$) of Andreev bound states, each characterized by a transmission coefficient, $T_i^{(k)}$ \cite{Beenakker:1991af}. The energy-phase relation of each JJ is then given by summing over the $i$ energies of the bound states,
\begin{equation}
	U_k (\varphi_k) = -\Delta \sum_{i} \sqrt{1-T_i^{(k)}\sin^2(\varphi_k/2)},
\end{equation}
where $\Delta$ is the superconducting gap and $\varphi_{k}$ is the superconducting phase difference across the $k^{\rm th}$ JJ. The total system Hamiltonian is given by
\begin{equation}
	H = 4E_C \hat{n}^2 - U_1(\hat{\varphi}) - U_2(\hat{\varphi}-2\pi\Phi/\Phi_0), \label{eqn:Hamiltonian}
\end{equation}
where $\Phi$ is the applied flux through the SQUID loop with $\Phi_0 = h/2e$ the superconducting flux quantum.  For identical, highly transmissive JJs at one-half flux quantum ($\Phi = \Phi_0/2$), odd harmonics in the Hamiltonian potential, $- U_1(\hat{\varphi}) - U_2(\hat{\varphi}-2\pi\Phi/\Phi_0)$, are suppressed, leaving a dominant $\cos(2\hat{\varphi})$ term and higher even harmonics. This results in a qubit with a $\pi$-periodic potential with coherent transport across the SQUID occurring only in units of $4e$ charge, that is, pairs of Cooper pairs. Here, the suppression of single-Cooper-pair transport results in the qubit having doubly degenerate ground states that differ by the parity of Cooper pairs on the island. Figs.~1(b) and (c) plot the qubit potential term at $\Phi = \Phi_0/2$ as a function of transmission coefficient $T_i^{(k)} = T$ (for all $i$, $k$) and symmetry parameter $\alpha = U_2(\hat{\varphi})/U_1(\hat{\varphi})$. Increasing asymmetry between the JJs increases coupling between the potential wells, resulting in a potential that is similar to that of a flux qubit. In the limit of strong asymmetry, $\alpha \to 0$, the single-well potential of a transmon qubit is recovered.

% Experimental setup
Figures 1(d)-(f) show micrographs of one of three measured devices. All devices showed similar spectra, with detailed time domain data taken on one of them. A large T-shaped island (blue) embedded in a ground plane was patterned from a 100 nm Al film on a high-resistivity silicon substrate, forming the shunting capacitor of the superconducting circuit. We estimate the charging energy of the island to be $E_C/h \sim 240$~MHz using electrostatic simulations. The semiconductor JJs are fabricated from molecular beam epitaxy-grown InAs nanowires with a $\sim$ 10~nm thick epitaxial aluminum layer grown on two of the nanowire facets. Each JJ is formed by etching away a $\sim$ 200~nm segment of the Al shell. The JJs are then connected between the island and the ground plane using evaporated Al contacts using \textit{in-situ} argon milling to remove native oxide layers. Proximal electrostatic gates (red) tune the JJ transmission by modulating the electron density predominantly in the junction region (green). The applied magnetic flux is controlled with the current through a nearby shorted transmission line while microwave excitations are driven using an open transmission line. The qubit is read out using a $\lambda/4$ cavity that is coupled with strength $g/2\pi \sim 80$~MHz to the qubit when operated in the transmon regime. The sample is measured in a dilution refrigerator at $<50$~mK inside superconducting Al and cryoperm magnetic shielding layers \cite{SuppInfo}. 

\begin{figure}
\includegraphics[width=1\columnwidth]{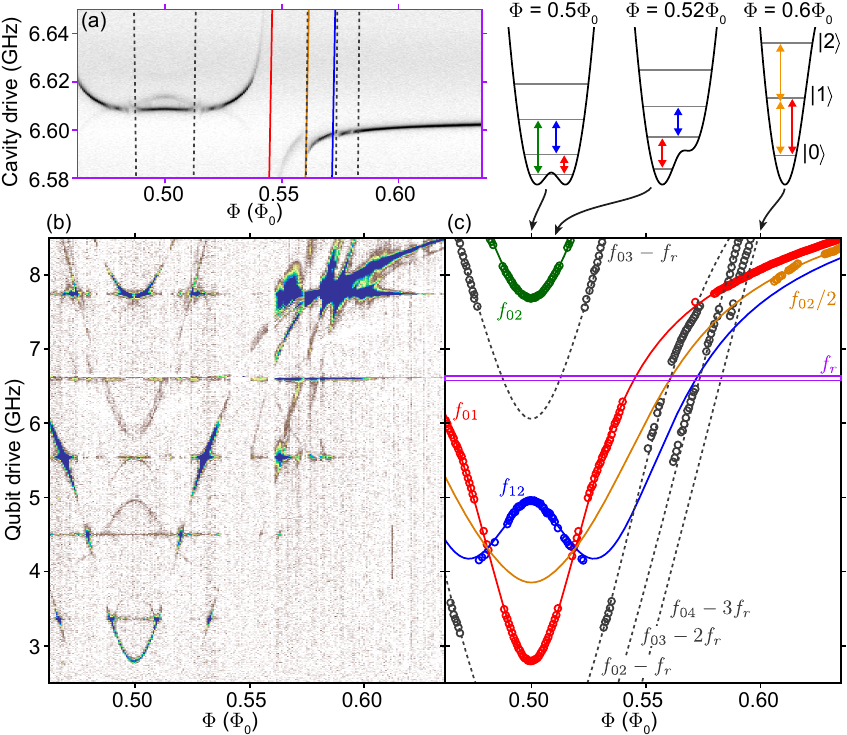}
\caption{\label{fig:2}
Qubit spectroscopy as a function of applied flux ($V_1 = 1.4$ V and $V_2 = -0.445$ V). (a) Frequency of the readout cavity as the qubit spectrum is tuned with flux. Solid and dashed lines indicate crossings of qubit transition frequencies as determined from fits to the data in (c). (b) Two-tone spectroscopy of the qubit transition frequencies. An average of each column has been subtracted. (c) Extracted transition frequencies from (b) with solid lines a fit to Eq. \eqref{eqn:Hamiltonian}. Diagrams above illustrate the fitted potential for different values of $\Phi$. Gray dashed lines indicate multi-photon transitions due to the simultaneous drive and readout tones.
}
\end{figure}

% Results (Fig. 2)
We first probe the readout cavity as a function of the flux through the SQUID, as shown in Fig.\ 2(a). Near one-half flux quantum a vacuum Rabi splitting is visible as the first excited cavity and qubit states hybridize (red line). Several other qubit states also weakly couple to the cavity, resulting in additional smaller anticrossings. We utilize two-tone spectroscopy to directly probe the transition frequencies of the qubit system: a readout tone, adjusted at each point in flux to the cavity frequency extracted from Fig.\ 2(a), was monitored while a second drive tone was swept in frequency to excite energy states. At a point tuned away from one-half flux quantum, we observe two transition frequencies with the spectrum resembling that of a transmon qubit with the higher frequency transition being $f_{01}$ (red) and a lower 2-photon excitation $f_{02}/2$ (orange). As the flux is tuned closer to one-half flux quantum,  the spectrum diverges from a transmon-like system, with anharmonicity, $h(f_{12}-f_{01})$, changing from negative to positive. Several horizontal lines are observed in the spectrum that we attribute to on-chip resonances, amplifying the readout response when coincident with a qubit transition frequency.

To understand the spectrum, we extract the excitation frequencies $f_{01}$, $f_{02}$, $f_{02}/2$, and $f_{12}$, shown as colored circles in Fig.\ 2(c). The extracted frequencies were fit by numerically calculating energy eigenstates of Eq.~\eqref{eqn:Hamiltonian}, taking $\Delta/h = 45$ GHz \cite{Chang:2015kw,SuppInfo} [solid lines in Fig.\ 2(c)]. From the fit we extract a charging energy $E_C/h = 284\pm 5$ MHz \cite{SuppInfo} and sets of transmission coefficients for each junction $\{T_i^{(1)}\} = \left\{1.0, 0.98, 0.29, 0.28\right\}$ and $\{T_i^{(2)}\} = \left\{0.95, 0.09, 0.09, 0.09\right\}$. Diagrams above Fig.\ 2(c) show the Josephson potential of the fitted model at different values of $\Phi$. At tuning condition $\Phi = \Phi_0/2$ the NW SQUID forms a symmetric double-well potential due to higher harmonics of the energy-phase relation. The barrier height between the two wells is tuned by the asymmetry of the two arms in the SQUID. Moving away from $\Phi=\Phi_0/2$, the potential is tilted, causing $f_{01}$ to sharply rise in energy, eventually resulting in a single well and the weakly anharmonic spectrum of the transmon. We match other transitions (gray dashed lines) to multi-photon excitations due to simultaneously applied readout and drive tones. These transition frequencies are calculated by subtracting an integer multiple of the cavity resonance frequency, $f_r$, from the fitted spectrum. Minor differences between the model and data may be due to small AC Stark shifts affecting the measured transition frequencies \cite{Schuster:2005df}.

\begin{figure}
\includegraphics[width=1\columnwidth]{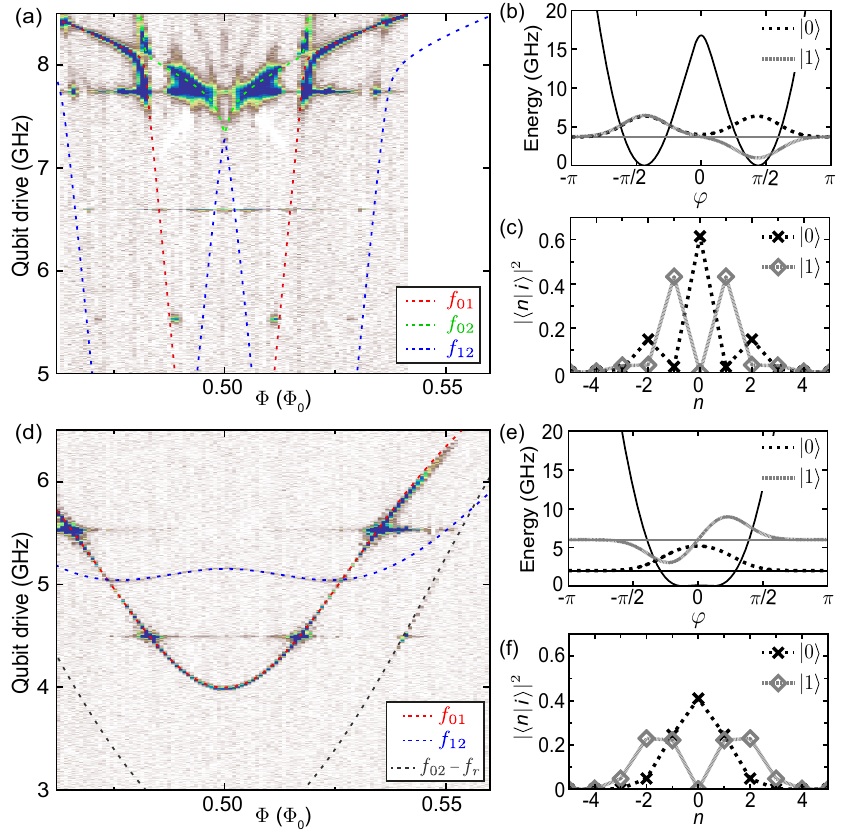}
\caption{\label{fig:3}
Voltage control of potential barrier. (a) Gate voltages are tuned to a balanced regime with the two junctions of similar value ($V_1 = 1.2$ V and $V_2 = -0.12$ V). (b) Qubit potential (solid black line) and wave functions for the two lowest energy states extracted from a fit to data in (a) at $\Phi = \Phi_0/2$. (c) The charge distribution of the two lowest energy states. (d) Gate voltages tuned to an unbalanced regime with one junction much smaller than the other ($V_1 = 1.241$ V and $V_2 = -0.386$ V). (e) Qubit potential and wave functions for the two lowest energy states extracted from fits to (d). (f) The charge distribution of the two states.
}
\end{figure}

% Results (Fig. 3)
Next, we study the effect of  modifying the gate voltages for each JJ. As highlighted in Fig.~1(c), the relative tuning of the two JJs can strongly modify the qubit potential. First, we tune into the protected qubit regime by adjusting the gate voltages to balance the two junctions such that single-Cooper-pair transport across the SQUID is suppressed, forming a double-well potential with minima separated by $\varphi\sim\pi$. In this balanced configuration, energy states are strongly localized to each of the wells with microwave-induced interwell transitions suppressed due to the small wave function overlap. The spectrum as a function of flux---controlling the tilt of the double-well potential---features transitions between the ground states and the next energy state of the same well [Fig.\ 3(a)]. Close to one-half flux quantum with a weakly tilted potential, $f_{01}$ is a forbidden transition between the two wells and is therefore not visible in the spectrum. As the potential is tilted further, two avoided crossings between $f_{01}$ and $f_{02}$ are observed when states $\ket{1}$ and $\ket{2}$, localized in separate wells, are on resonance. The spectrum is reminiscent of a heavy fluxonium, which also has a double potential well but with minima separated by $2\pi$ instead of $\pi$ \cite{Lin:2017uz,Earnest:2017tf}. As for Fig.~2, we extract the transition frequencies and fit them to Eq.~\eqref{eqn:Hamiltonian} with $E_C/h = 284$ MHz and $\Delta/h = 45$ GHz to find the transmission coefficients $\{T_i^{(1)}\} = \left\{1.0, 1.0, 0.60, 0.0, 0.0\right\}$ and $\{T_i^{(2)}\} = \left\{0.99, 0.78, 0.31, 0.30\right\}$. At $\Phi = \Phi_0/2$ the potential forms a double-well potential with minima at $\varphi\sim\pm \pi/2$ with two nearly degenerate ground states given by the bonding and anti-bonding eigenstates [Fig.\ 3(b)]. In Fig.~3(c), the two ground states are plotted in the charge basis, clearly showing the separation in parity with either odd or even numbers of Cooper pairs.

\begin{figure}
\includegraphics[width=1\columnwidth]{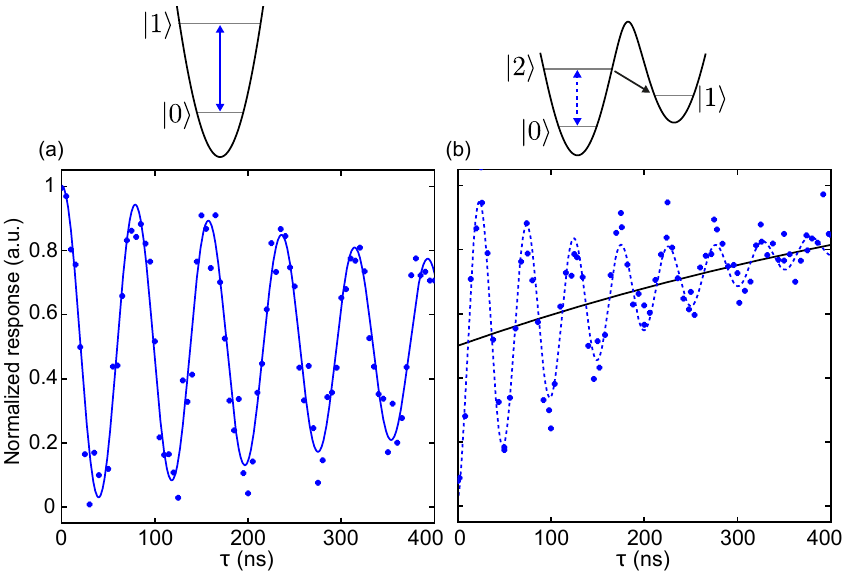}
\caption{\label{fig:4}
Coherent control. Rabi oscillations in (a) the transmon regime with $\Phi = 0$ and $f_\text{Drive} = 7.911$ GHz and (b) a tilted $\pi$ qubit regime with $\Phi = 0.512\Phi_0$ and $f_\text{Drive} = 5.725$ GHz. Diagrams show the qubit potentials and lowest energy states. Voltages are fixed at $V_1 = -1.25$ V and $V_2 = -0.445$ V for both (a) and (b). The solid line in (a) is a fit to an exponentially decaying sinusoidal function, while in (b) the fit function has an additional exponentially decaying offset (black line).
}
\end{figure}

In {color{red} contrast}, Fig.~3(d) shows the qubit spectrum with one junction tuned to have much lower total transmission than the other. Again, fitting the measured spectrum to Eq. \eqref{eqn:Hamiltonian} yields $\{T_i^{(1)}\} = \left\{1.0, 0.91, 0.30, 0.20, 0.18\right\}$ and $\{T_i^{(2)}\} = \left\{0.90, 0.06, 0.06, 0.06\right\}$. The potential and two lowest energy states extracted from the fit [Fig.\ 3(e) and (f)] has the form of a harmonic oscillator with a small perturbation giving a positive anharmonicity, similar to a flux qubit \cite{Yan:2016df}. 

\begin{figure}
\includegraphics[width=1\columnwidth]{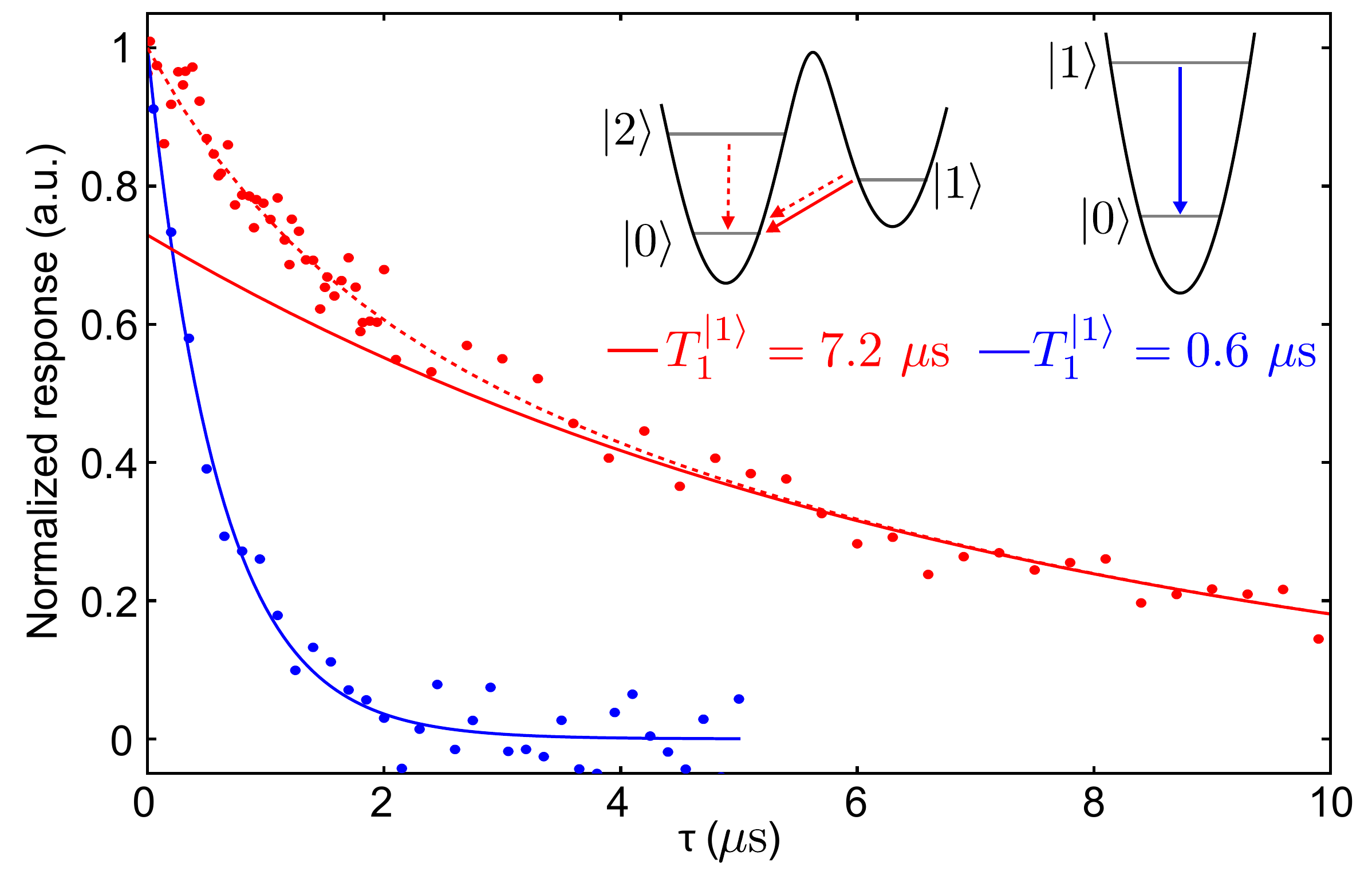}
\caption{\label{fig:5}
Lifetime measurements in the transmon regime with $\Phi = 0$ and $f_\text{Drive} = 7.911$ GHz (blue) and the tilted potential regime with $\Phi = 0.512\Phi_0$ and $f_\text{Drive} = 5.725$ GHz (red). Blue solid line is an exponential fit. Red dashed line is a double exponential fit $A_1e^{-\tau/T_1^{|1\rangle}}+A_2e^{-\tau/T_1^{|2\rangle}}$ with the solid line showing $A_1e^{-\tau/T_1^{|1\rangle}}$ (see main text). Data are normalized to fit parameters.
}
\end{figure}

% Results (Fig. 4)
Using {\it in-situ} gate control, we are able to demonstrate protection of coherence in the symmetric (balanced) regime compared to the asymmetric (transmon) regime. Due to the protection of the 4$e$-parity states of the qubit, states are indistinguishable through dispersive measurements. Instead, we operated at $\Phi = 0.512\Phi_0$ giving a slightly tilted double well potential while allowing visible readout [Fig.\ 4(b) diagram]. In the asymmetric (transmon) regime, applying a drive tone at the qubit resonance frequency for a time $\tau$ yielded Rabi oscillations, as shown in Fig.\ 4(a). On the other hand, near the symmetric configuration, the $|0\rangle\leftrightarrow|1\rangle$ transition was strongly forbidden, prevent a direct comparison to the asymmetric regime. Instead, in the near-symmetric case, we instead applied microwave drive at the unprotected $|0\rangle-|2\rangle$ transition frequency. Figure 4(b) shows the microwave-induced Rabi oscillations between the $|0\rangle$ and $|2\rangle$ states in this configuration.  Oscillations occur around an exponentially decaying offset (black line), which we interpret as decay from the $|2\rangle$ state to the $|1\rangle$ state, trapping the population in $|1\rangle$ at long drive times.

% Results (Fig. 5)
We measured qubit lifetime in the unprotected regime by applying a $\pi$-pulse followed by readout after a wait time $\tau$ [Fig.~5, blue]. Fitting the data to an exponential decay, we extract a lifetime $T_1 = 0.6~\mu$s. Near the protected regime, we drove the $|0\rangle\leftrightarrow|2\rangle$ transition with a long $3~\mu$s pulse to initialize the $|1\rangle$ state, followed by readout after a wait time $\tau$ [Fig.\ 5, red]. We observe two superimposed exponential decays with lifetimes $T_1^{|1\rangle} = 7.2~\mu$s and $T_1^{|2\rangle} = 1.2~\mu$s that we interpret as relaxation from an incoherent mixture of the $|1\rangle$ and residually populated $|2\rangle$ states, respectively. The factor of $\sim12$ enhancement in $|1\rangle$ state lifetimes near the protected regime is qualitatively consistent with a suppressed charge matrix element, $\left\langle 0\right|\hat{n}\left|1\right\rangle \rightarrow 0$. Extracted matrix elements indicate much longer lifetimes are achievable \cite{SuppInfo}, and we speculate that lifetimes becomes limited by decay channels such as a residual resistance of the semiconductor JJs due to subgap states \cite{Antipov:2018wz,Vaitie:2018aa}.

In summary, we have demonstrated a superconducting circuit architecture based on tunable, high-transmission semiconductor JJs configured to realize a parity-protected qubit. The simplicity and \textit{in-situ} tunability of this circuit along with recently reported semiconductor two-dimensional-electron-gas-based JJs \cite{Casparis:2018aa} paves the way for scalable, parity-protected qubit. Furthermore, we have demonstrated dispersive readout of qubit states with enhanced lifetimes by operating with a small detuning from the protected regime. This points to readout of protected states by dynamically modifying the device tuning to lift the degree of protection. Alternatively, protected qubit states might be distinguished using parametrically driven readout schemes \cite{Didier:2015ej}. Finally, further work might take advantage of recently demonstrated high-impedance resonators ($Z\gg 1~\text{k}\Omega$) \cite{Hazard:2019aa,Niepce:2019aa,Zhang:2019aa} and fast superconducting switches such as superconducting FETs \cite{Casparis:2019aa} to implement a topological qubit with protected qubit operations \cite{Brooks:2013hi}.

%Acknowledgements
\begin{acknowledgements}
We thank Lev B. Ioffe for valuable discussions. This work was supported by Microsoft, the U.S. Army Research Office, and the Danish National Research Foundation. N.P. acknowledges support from the Swiss National Science Foundation and NCCR QSIT, F.K. acknowledges support from the Danish Innovation Fund and C.M.M. acknowledges support from the Villum Foundation.
\end{acknowledgements}

\bibliography{Bibtex}

\onecolumngrid
\clearpage
\onecolumngrid
\setcounter{figure}{0}
\setcounter{equation}{0}
\section{\large{S\texorpdfstring{\MakeLowercase{upplemental} M\MakeLowercase{aterial}}{}}}
\renewcommand\thefigure{S\arabic{figure}}
\renewcommand{\tablename}{Table.~S}
\renewcommand{\thetable}{\arabic{table}}
\renewcommand{\theequation}{S\arabic{equation}} 
\renewcommand{\ket}[1]{{\left\vert #1\right\rangle}}
\renewcommand{\bra}[1]{{\left\langle #1\right\vert}}

%\twocolumngrid

\vspace{2cm}
Contents
\begin{enumerate}
\item Numerical simulation of eigenstates and fitting of energy spectra
\item Energy spectrum and matrix elements
\item Experimental setup
\end{enumerate}

\subsection{1. Numerical simulation of eigenstates and fitting of energy spectra}

We fit the data to the Hamiltonian of two high-transmission Josephson junctions in a SQUID geometry given by:
\begin{equation}
	H = 4E_C \hat{n}^2 -  \sum_{i}\Delta \sqrt{1-T_i^{(1)}\sin^2(\hat{\varphi}/2)} - \sum_{i}\Delta\sqrt{1-T_i^{(2)}\sin^2[(\hat{\varphi}-\phi)/2]},
\end{equation}
where $\phi = 2\pi\Phi/\Phi_0$. 
For numerical simulations of the Hamiltonian, we first rewrite the Hamiltonian in charge basis.
The Josephson junctions given in phase basis is transformed into charge basis by performing a discrete Fourier transform of the energy-phase relation:
\begin{align}
	-  \sum_{i} \Delta\sqrt{1-T_i\sin^2[(\hat{\varphi}-\phi)/2]} =& - \sum_{k=1}^{\infty} E_k\cos[k(\hat{\varphi}-\phi)] \\
	=& - \sum_{k=1}^{\infty} E_k\frac{e^{ik(\hat{\varphi}-\phi)}+e^{-ik(\hat{\varphi}-\phi)}}{2} \\
	=& - \sum_{k=1}^{\infty} E_k\frac{e^{-ik\phi}|n\rangle \langle n+k|+e^{ik\phi}|n+k\rangle \langle n|}{2}.
\end{align}
The Hamiltonian can then be written in the charge basis as:
\begin{align}
	H = 4E_C n^2 \ket{n}\bra{n} &-  \sum_{k=1}^{\infty} E_k^{(1)}\frac{|n\rangle \langle n+k|+|n+k\rangle \langle n|}{2} \nonumber \\
	&- \sum_{k=1}^{\infty} E_k^{(2)}\frac{e^{-ik\phi}|n\rangle \langle n+k|+e^{ik\phi}|n+k\rangle \langle n|}{2} \nonumber \\
	= 4E_C n^2 \ket{n}\bra{n} &-  \sum_{k=1}^{\infty}\frac{E_k^{(1)} + e^{-ik\phi}E_k^{(2)}}{2}|n\rangle \langle n+k|+\frac{E_k^{(1)} + e^{ik\phi}E_k^{(2)}}{2}|n+k\rangle \langle n|,
\end{align}
where $E^{(1)}_k(\mathcal{T}^{(1)},\Delta)$ and $E^{(2)}_k(\mathcal{T}^{(2)},\Delta)$ are the Fourier components of each Josephson junction.
In a basis of charge states, $\ket{n}$, given by  $\mathcal{V}=\left\{\dots,\ket{2},\ket{1},\ket{0}, \ket{-1},\dots\right\}$, the Hamiltonian is the matrix:
\begin{equation}
H = \begin{bmatrix}
	\ddots & \vdots & \vdots & \vdots & \vdots & \reflectbox{$\ddots$} \\
    \dots & 16E_C & \frac{-E_1^{(1)} - e^{i\phi}E_1^{(2)}}{2} & \frac{-E_2^{(1)} - e^{i2\phi}E_2^{(2)}}{2} & \frac{-E_3^{(1)} - e^{i3\phi}E_3^{(2)}}{2} & \dots \\
    \dots & \frac{-E_1^{(1)} - e^{-i\phi}E_1^{(2)}}{2} & 4E_C & \frac{-E_1^{(1)} - e^{i\phi}E_1^{(2)}}{2} & \frac{-E_2^{(1)} - e^{i2\phi}E_2^{(2)}}{2} & \dots \\
    \dots & \frac{-E_2^{(1)} - e^{-i2\phi}E_2^{(2)}}{2} & \frac{-E_1^{(1)} - e^{-i\phi}E_1^{(2)}}{2} & 0 & \frac{-E_1^{(1)} - e^{i\phi}E_1^{(2)}}{2} & \dots \\
    \dots & \frac{-E_3^{(1)} - e^{-i3\phi}E_3^{(2)}}{2} & \frac{-E_2^{(1)} - e^{-i2\phi}E_2^{(2)}}{2} & \frac{-E_1^{(1)} - e^{-i\phi}E_1^{(2)}}{2} & 4E_C & \dots \\
    \reflectbox{$\ddots$} & \vdots & \vdots & \vdots & \vdots & \ddots
\end{bmatrix}.
\end{equation}
For numerical simulations, we truncate the Hilbert space at 41 states ($\mathcal{V}=\left\{\ket{20},\dots,\ket{1},\ket{0}, \ket{-1},\dots,\ket{-20}\right\}$) and set $E_k = 0$ if $E_k < 1$~MHz. 
For a given value of $\phi$, eigenenergies $E_\ket{i}$ and eigenvectors $\psi_{\ket{i}}(n)$, where $\ket{i}$ refers to the $i$'th eigenstate of the matrix, are found numerically with \texttt{numpy.linalg.eig()} in \texttt{Python}. 
Transition frequencies of the model at $\phi$ are readily calculated as the energy differences of the sorted set of eigenenergies ($f_{01} = (E_\ket{1}- E_\ket{0})/h$).
Eigenvectors of the matrix are wavefunctions of quantum states in charge basis as plotted in Figure 3 of the main text. 
The wavefunctions in phase basis are calculated from the relation $\psi_\ket{i}(\varphi) = \sum_n e^{in\varphi}\psi_\ket{i}(n)$. 
The charge matrix elements are computed as $\bra{k}n\ket{i}=\sum_{n=-20}^{20}\psi_\ket{k}(n)^\dagger n \psi_\ket{i}(n)$.

To fit the data we use \texttt{scipy.optimize.least\_squares()} to find the sets of transmissions $\{T_i^{(1)}\}$ and $\{T_i^{(2)}\}$ ($\Delta$ is fixed) that minimizes the differences between numerically simulated transition frequencies and measured transition frequencies for all measured values of $\phi$.

The fit procedure finds good agreement with measurement data, varying only the junction transmission coefficients to account for different device tunings.
However, we find a discrepancy between the charging energy, $E_C/h$, estimated from electrostatic simulations and from fits to the data of 235 MHz and 280 MHz respectively.
This could be due to the assumption of fixed gap energy, $\Delta$, for all channels in both junctions or simplifications of the model such as not accounting for charge renormalization due to the transmission coefficients approaching unity \cite{Averin:1999uo,bargerbos:2019aa,kringhoj:2019xa}.

\subsection*{2. Energy spectrum and matrix elements}
\begin{figure}[h]
\centering
\includegraphics[width=0.8\hsize]{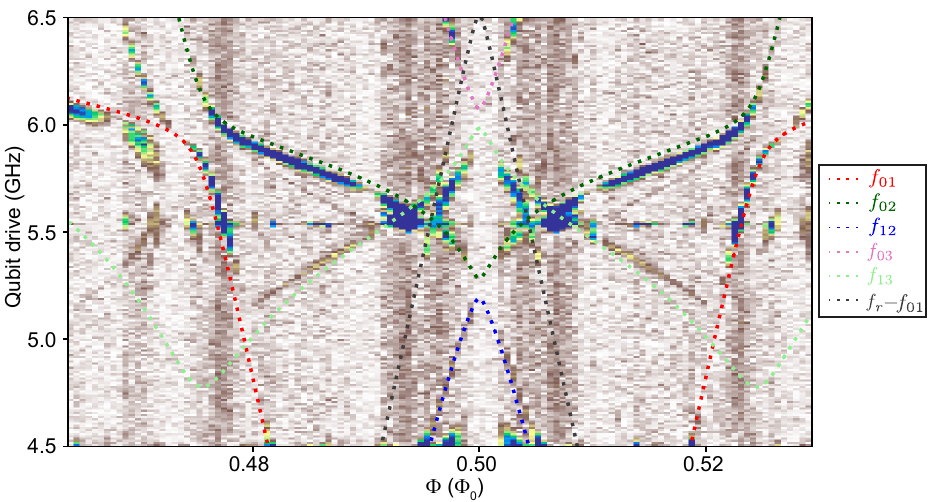}
\caption[Energy spectrum]{\label{FigS:1}
Energy spectrum for gate voltages at $V_1 = -1.25$~V and $V_2 = -0.445$~V.
}
\end{figure}

\begin{figure}[h]
\centering
\includegraphics[width=0.8\hsize]{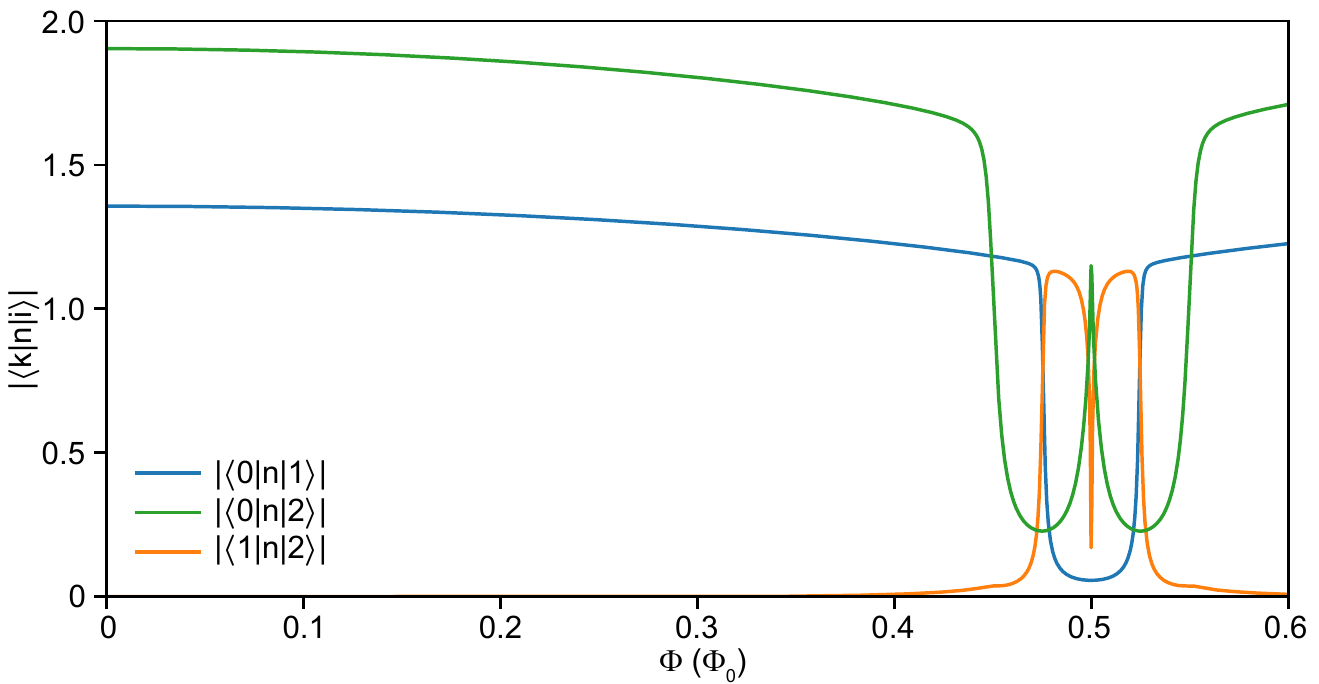}
\caption[Charge matrix elements]{\label{FigS:2}
Charge matrix elements for gate voltages at $V_1 = -1.25$~V and $V_2 = -0.445$~V.
}
\end{figure}

Figure \ref{FigS:1} shows spectroscopy data used to extract potentials plotted in Figures 4 and 5 of main text.
Figure \ref{FigS:2} shows calculated charge matrix elements for the fitted model.

\subsection{3. Experimental setup}
\begin{figure}
\includegraphics[width=1\hsize]{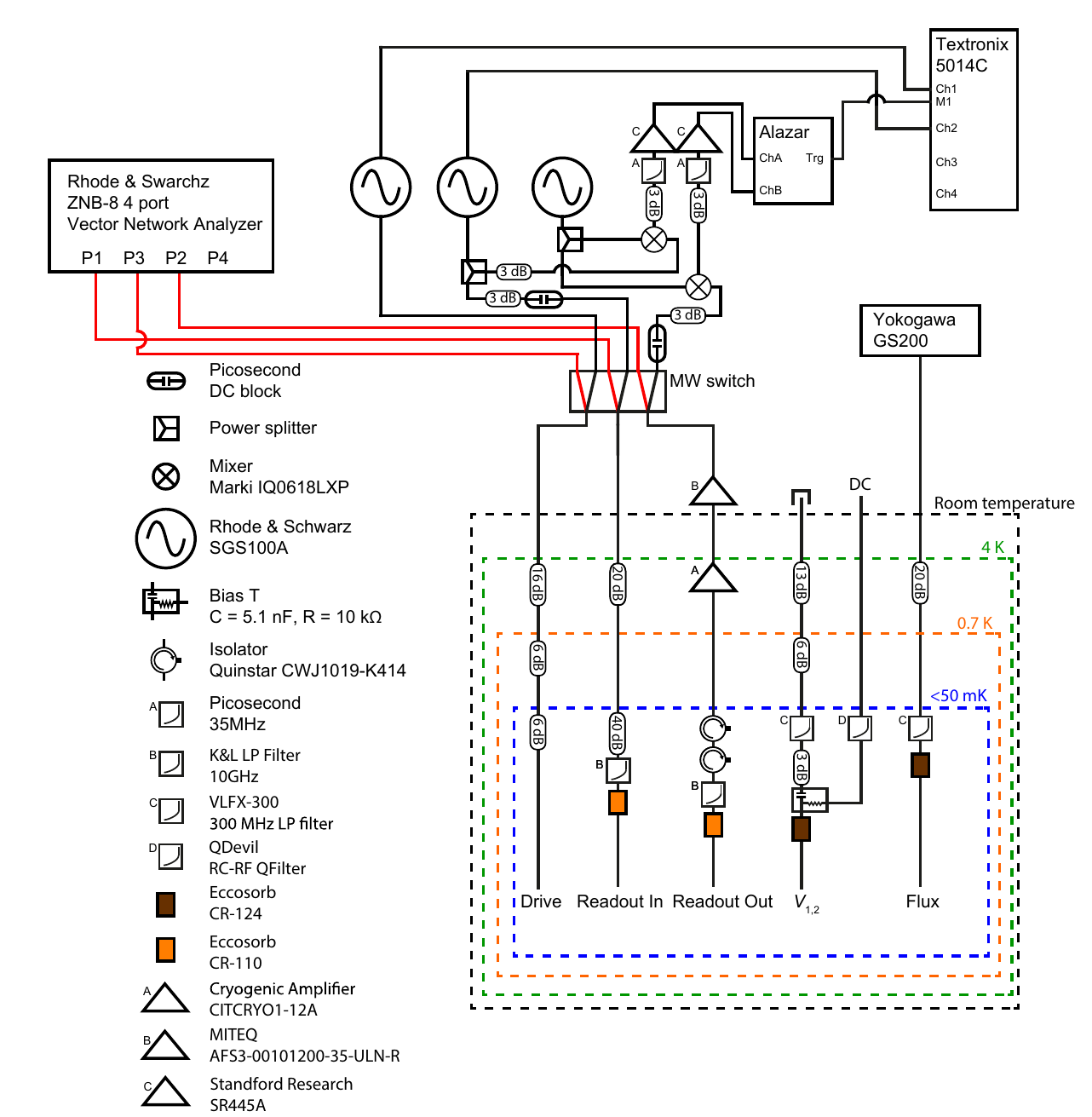}
\caption{\label{FigS:3}
Experimental setup.
}
\end{figure}

The experimental setup is presented in Fig. \ref{FigS:3}. For spectroscopy data in figures 2 and 3 of main text the microwave switch was connected to the VNA through red lines. Time domain data in figures 4 and 5 was taken witch switch connected to black lines. 

\end{document}